\documentclass[3p,times,twocolumn]{elsarticle}

\usepackage{ecrc}

\volume{00}

\firstpage{1}

\journalname{Nuclear Physics B Proceedings Supplement}

\runauth{S. Mangano}

\jid{nuphbp}

\jnltitlelogo{Nuclear Physics B Proceedings Supplement}

\usepackage{amssymb}

\usepackage[figuresright]{rotating}

\begin{document}

\begin{frontmatter}



\dochead{}

\title{Astrophysical point source search with the ANTARES neutrino telescope}


\author{S. Mangano for the ANTARES Collaboration}

\address{IFIC - Instituto de F\'isica Corpuscular, Edificio Institutos de Investigaci\'on, Apartado de Correos 22085, 46071, Spain}

\begin{abstract}
\vspace*{-0.2cm}
The ANTARES neutrino telescope is 
installed at a depth of 2.5 km of the
Mediterranean Sea and consists of
a three-dimensional array of 885 photomultipliers 
arranged on twelve detector lines.
The prime objective is
to detect high-energy neutrinos from extraterrestrial origin.
Relativistic muons emerging from charged-current muon neutrino
interactions in the detector surroundings produce a cone of Cerenkov light
which allows the reconstruction of the original neutrino direction.
The collaboration has implemented different methods to
search for neutrino point sources in the data collected since 2007.
Results obtained with these methods as well as the sensitivity of the
telescope are presented.
\end{abstract}


\end{frontmatter}


\vspace*{-1.55cm}
\section{Algorithms to search point sources}
\vspace*{-0.2cm}
\label{}
The following two unbinned clustering algorithms are used to 
search for neutrino point sources (signal) in the sky map of atmospheric 
neutrinos (background). 
The \textit{Expectation-Maximization} algorithm is based 
on an analytically likelihood maximization~\cite{ag}. The signal density 
distribution is assumed to be a two-dimensional gaussian and 
the background density distribution is taken from 
the data. The probability to have a signal 
for a given background model is maximized. The free parameters are the 
two sigmas of the two-dimensional gaussian probability 
density function and the expected number of signal events from the source.

The \textit{Likelihood algorithm} calculates first the 
angular distance between a fixed source search point and 
the location of all selected neutrinos in the sky. Then it fits 
this distribution with signal and background density distribution 
using maximization technique. The signal distribution is the 
point spread function taken from Monte Carlo 
and the background distribution is taken from data.

\vspace*{-0.25cm}
\section{Data results}
\begin{figure}[h]
\centering
\vspace*{-0.75cm}
\includegraphics[width=0.46\textwidth,height=0.45\textwidth,angle=0]{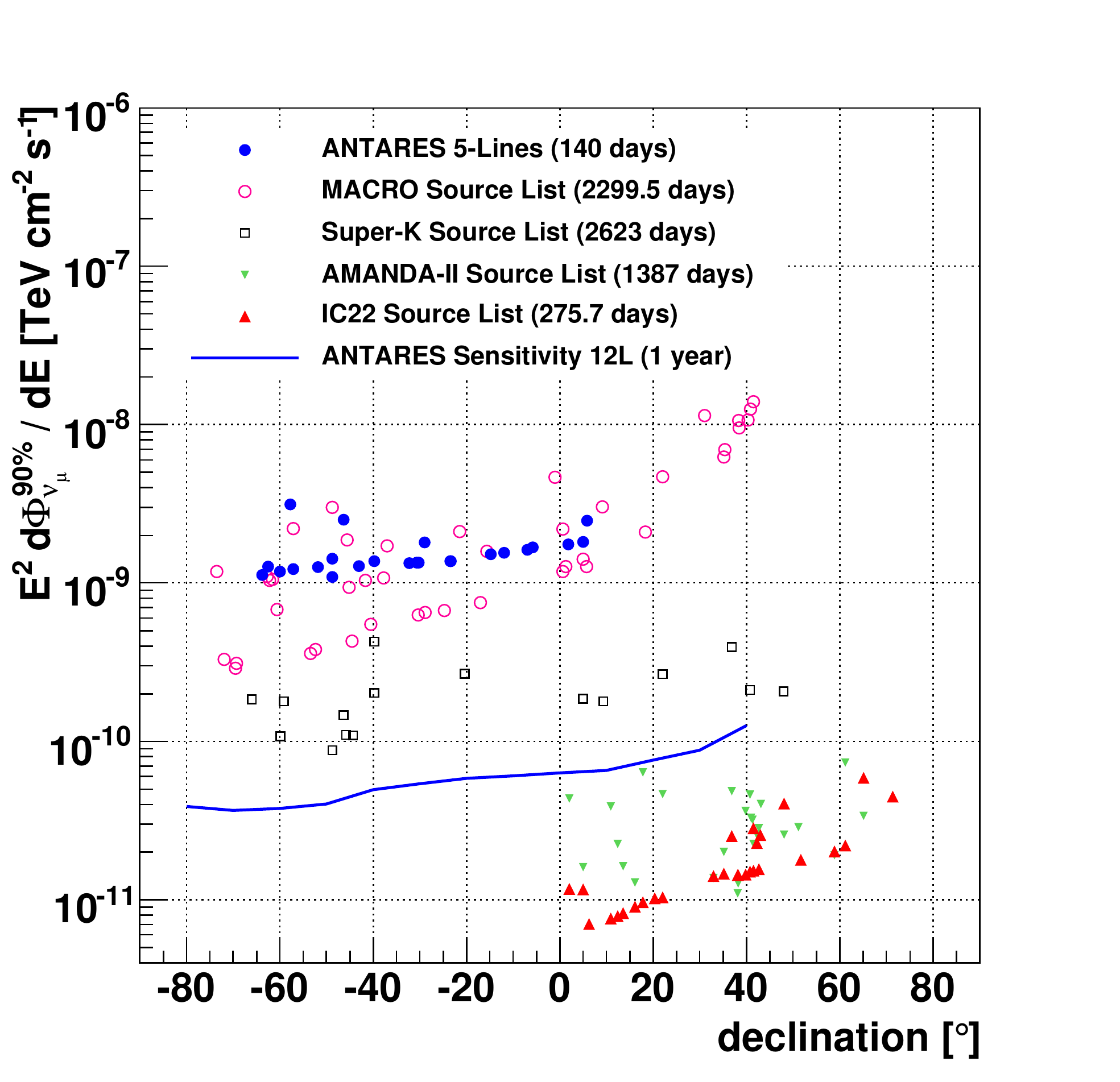}
\vspace*{-0.5cm}
\caption[Sc]{\textit{ The upper limits from the 24 candidate sources for 2007 data (filled points) compared with the results published by other neutrino experiments (Macro~\cite{am}, Super-K~\cite{de}, Amanda~\cite{ab} and IceCube~\cite{ic}). The predicted sensitivity of ANTARES for 365 days (line) is also shown.}}
\label{fig:plot}
\end{figure}

\vspace*{-0.2cm}
A set of cuts have been optimized to search for an $E^{-2}$ 
neutrino flux in the data sample taken in 2007 with a livetime of 140 days. 
An all sky point source search based on the above algorithms 
has not revealed any significant excess for any direction. No 
significant excess has been found also in 
a dedicated search from a selected list 
of 24 promising neutrino source candidates. 
The upper limits on the neutrino flux from the 
24 candidate sources are shown in Figure \ref{fig:plot}.

\vspace*{-0.25cm}
\section*{Acknowledgments}
\vspace*{-0.2cm}
\scriptsize
I gratefully acknowledge the support of the JAE-Doc postdoctoral programme of CSIC. 
This work has also been supported by the following 
Spanish projects: FPA2009-13983-C02-01, MultiDark Consolider CSD2009-00064, ACI2009-1020 of MICINN and Prometeo/2009/026 of Generalitat Valenciana.












\end{document}